\documentclass[sigplan,10pt]{acmart}
\renewcommand\footnotetextcopyrightpermission[1]{}
\usepackage{filecontents}
\usepackage{graphicx}
\usepackage{enumitem}
\usepackage{pifont}

\begin{document}

\date{}

\title{ Varuna: Scalable, Low-cost Training of \\
Massive Deep Learning Models}
\author{Sanjith Athlur}
\affiliation{Microsoft Research India}
\authornote{Equal Contribution}
\author{Nitika Saran}
\affiliation{Microsoft Research India}
\authornotemark[1]
\author{Muthian Sivathanu}
\affiliation{Microsoft Research India}
\author{Ramachandran Ramjee}
\affiliation{Microsoft Research India}
\author{Nipun Kwatra}
\affiliation{Microsoft Research India}

\newcommand{\eg}{\textit{e.g.}}
\newcommand{\ie}{\textit{i.e.}}
\newcommand{\etal}{\textit{et al.}}
\newcommand{\etc}{\textit{etc.}}
\newcommand{\adhoc}{\textit{ad hoc}}

\newcommand{\varuna}{\textit{Varuna}}
\newcommand{\msft}{\textit{Company-X}}

\begin{abstract}
    Systems for training massive deep learning models (billions of parameters) today assume and require specialized "hyper-clusters": hundreds or thousands of GPUs wired with specialized high-bandwidth interconnects such as NV-Link and Infiniband. Besides being expensive, such dependence on hyper-clusters and custom high-speed inter-connects limits the size of such clusters, creating (a) scalability limits on job parallelism; (b) resource fragmentation across hyper-clusters.

In this paper, we present \varuna\, a new system that enables training massive deep learning models on commodity networking.  \varuna\ makes thrifty use of networking resources and automatically configures the user’s training job to efficiently use any given set of resources. Therefore, \varuna\ is able to leverage “low-priority" VMs that cost about 5x cheaper than dedicated GPUs, thus significantly reducing the cost of training massive models. We demonstrate the efficacy of \varuna\ by training massive models, including a 200 billion parameter model, on 5x cheaper “spot VMs", while maintaining high training throughput. 
Varuna improves end-to-end training time by up to 18x compared to other model-parallel approaches and up to 26\% compared to other pipeline parallel approaches

The code for \varuna\ is available at \url{https://github.com/microsoft/varuna}.
\end{abstract}

\settopmatter{printfolios=true}
\maketitle
\pagestyle{plain}

\section{Introduction}

State-of-the-art deep learning models that power important applications such as web search, have seen a rapid growth in number of model parameters.   For example, in natural language understanding,  BERT-large~\cite{bert-large} enabled significant accuracy improvements with a model architecture that has 340 million parameters, significantly larger than any other model at that time.  Subsequently, GPT-2~\cite{gpt-2} (1.5 billion parameters), Megatron~\cite{shoeybi2019megatron} (8 billion parameters), Turing-NLG~\cite{Turing} (17 billion parameters), and GPT-3~\cite{gpt-3} (175 billion parameters) have pushed model sizes much higher, improving accuracy even further in the process.

Such massive models require several petaflops of compute, so they need to be run on large number of GPUs. Because the training is also communication-intensive, such massive jobs are run on specialized ``hyperclusters" that have expensive, high-speed interconnects such as NVLink or Infiniband. Even in these hyperclusters, Megatron 8B model takes roughly 11 days to train on 512 GPUs~\cite{shoeybi2019megatron}.
Thus, training such large models can be expensive. Authors in~\cite{trainingcost} estimate that fully-loaded training cost, which includes multiple training runs with hyper-parameter tuning, for BERT-large to be \$200K.

Besides cost, the dependence on specialized hyperclusters for training massive models is also sub-optimal for other reasons.  First, specialized high-speed interconnects are economically infeasible to scale beyond a certain cluster size.  For example, Nvidia's NVLink connects 16 GPUs within a DGX-2 server at 2.4~Tbps all-to-all bandwidth but the DGX-2 servers themselves are connected with each other using Infiniband at only 800~Gbps~\cite{nvlink-dgx2}. 
Thus, the degree of parallelism for training massive models is limited by the architecture and size of a single hypercluster (\eg, 1000 GPUs).   Second, multiple siloed hyper-clusters cause resource fragmentation, as GPUs are not fungible across hyper-clusters.

In this paper, we present \varuna, a system that trains massive deep learning models on commodity networking, without requiring specialized hyperclusters, thus addressing the three problems with existing approaches: cost, scale, and resource utilization.  As \varuna\ does not depend on specialized networking, it can orchestrate a job in any set of GPUs in the data center, thus improving resource utilization.

Interestingly, the ability of \varuna\ to harness scattered GPUs in the data center makes it possible to train massive jobs on ``low-priority" or ``spot" VMs that public clouds offer~\cite{azure-lowprivm,aws-spotvm}.   Such spot VMs are offered at a significant discount (\eg, 4-5x cheaper) compared to dedicated VMs, as they allow the cloud provider to sell unused spare capacity, and manage capacity better during load spikes.  By opening up such cheap VMs for training massive models that today run on hyperclusters, \varuna\ reduces the cost of training such models by 4-5x, without losing performance.

There are three key challenges that \varuna\ addresses: (a) dealing with ``slow''/flaky network,  (b) dealing with transient pre-emptible resources, and (c) being transparent/non-intrusive to the programmer.

First, \varuna\ handles low-bandwidth networks with a new variant of {\em pipeline parallelism} combined with data parallelism within each pipeline stage to train massive models.   We show that {inter-layer} or pipeline-partitioning (\eg, GPipe~\cite{Gpipe}, Pipedream~\cite{pipedream}) is more tolerant of slow networks compared to the more popular {\em intra-layer} partitioning (employed by Mesh-Tensorflow~\cite{meshtf}, Megatron~\cite{shoeybi2019megatron}, Turing-NLG~\cite{Turing}).  To improve pipeline efficiency, \varuna\ uses a large number of {\em micro-batches} within a mini-batch, similar to Gpipe, but enhanced with a {\it novel, micro-batch scheduling algorithm} that is more efficient and tolerant to network jitter.   Further, unlike traditional pipeline partitioning that tries to minimize the number of partitions (\ie, each partition fits as much work as possible, constrained only by GPU memory), stages in \varuna\ are also constrained by the network bandwidth.  As the synchronization of gradients for data parallelism happens only within a partition, \varuna\  limits bandwidth usage by scaling the number of pipeline stages as model size increases. 
One issue with large number of micro-batches is that it increases mini-batch size, raising model accuracy concerns; we address it by demonstrating, for the first time, that a large 2.5 billion-parameter GPT-2 model can be trained to the same accuracy, despite using a 16x larger mini-batch.

Second, \varuna\ handles frequent pre-emptions of low-priority VMs by employing dynamic, semantics-preserving reconfiguration of the training job.  Existing approaches for elasticity like  PyTorch Elastic~\cite{elastic} and MXNet Dynamic~\cite{mxnetdl} require users to provide different sets of hyper-parameters and mini-batch sizes for a varying number of resources.  Recent work~\cite{or2020resource} allows the user to specify a single mini-batch size, but suffers poor performance at high scaling factors. Crucially, all the above approaches only handle data parallelism.  Instead, \varuna\ introduces {\em job morphing}, where it morphs the configuration of a large parallel job  across both the pipeline depth and data parallelism dimensions, to fit in as much resources as available, {\it without changing hyper parameters}.  \varuna\ piggybacks on the gradient accumulation that micro-batching performs, adapting to a wide range of ``local'' batch-sizes.   For such morphing, \varuna\ uses a novel mechanism of micro-benchmark-driven simulation, where a small set of micro-parameters are calibrated through profiling, that are then fed to an efficient simulator, to pick the best performing configuration for different number of GPUs.
\varuna\ also deals with failures and stragglers, inherent to the spot-VM setup.

Third, \varuna\ addresses a key usability challenge of model partitioning, by making it non-intrusive on the programmer, unlike existing approaches that require significant changes to the model~\cite{Gpipe,zero}.  \varuna\ provides flexibility to the user in writing the model as if it runs on a single GPU.   To achieve this, \varuna\ introduces a novel mechanism of automatically identifying {\em cut-points} in the model,  which are a {\em super-set} of potential ``safe'' partitioning points.  \varuna\ then groups multiple of these cut-points into a single partition.  Another challenge is implicit usage of cross-partition state (\eg, global computations in optimizers like NVLAMB~\cite{NVLAMB}, shared weights, \etc.), either directly in the user's code or indirectly in libraries (such as APEX~\cite{APEX}) used by the model; \varuna\ includes a tracer that identifies such cross-partition sharing automatically across libraries, and flags those tensors to be synchronized across partitions.

We evaluate \varuna\ by using it to train several massive models: BERT-large (340 million parameters), GPT-2 models with various sizes: 2.5 billion, 8.3 billion, 20 billion, and 200 billion.     All these models are trained on {\em low-priority VMs} in Azure that cost 4-5x cheaper than their dedicated counterparts. On such commodity VMs, we show that \varuna\ improves the performance of training such models by up to 18x compared to state-of-the-art approaches.   We also show that despite the commodity networking across these VMs, \varuna\ is able to {\em outperform} state-of-the-art approaches that run on specialized hyperclusters by upto 18x. 

We make the following key contributions in this paper:

\begin{itemize}[nosep]

\item
We challenge the pervasive belief (and practice) that massive models can be trained only on specialized hyperclusters, by presenting the first system that is capable of training massive deep learning models on spot VMs with commodity networking, achieving 4-5x lower cost of training these models.

\item
We argue and demonstrate that intra-layer partitioning is ill-suited not only for commodity networking but that they are {\it not the best performing option even in hyperclusters}.

\item
We introduce a novel concept of correctness-preserving {\em job morphing} to automatically reconfigure a running DLT job, to adapt to changing number of GPUs, using a combination of data and model parallelism.

\item
We demonstrate the efficacy of this approach by efficiently scaling to a 200 billion parameter model, and showing significant speedups on other large models such as BERT-large and Megatron-8.3B.  We also  demonstrate that despite using a large batch size, \varuna\ achieves state-of-the-art accuracy on a 2.5 billion parameter GPT-2 model.
\end{itemize}

The code for {\it Varuna} has been open-sourced  and is available at \url{https://github.com/microsoft/varuna}.
\section{Background and Related Work}
\label{sec-background}

In this section, we present a brief background of Deep Learning Training jobs (DLT jobs) and review today's architectures for training massive models.

A DLT job is typically a python script that uses frameworks such as PyTorch~\cite{pytorch} or TensorFlow~\cite{tensorflow} to define the model and the training procedure.  In each training iteration, the DLT job takes a few samples of data called the {\it mini-batch} as input and performs a {\it forward pass} over this data. The forward pass consists of applying the model function on the input data to compute a loss value. The loss value is then propagated in a {\it backward pass} that computes a {\it gradient} for each model parameter. Finally, the model parameters are updated by adding the negative of the gradient, scaled by a hyper-parameter called the {\it learning rate}. Millions of such iterations are typically necessary to learn an accurate model.  Since the forward and backward passes involve billions of floating point operations, they are performed in a GPU.  The model parameters as well as gradients remain in GPU memory during training for maximum efficiency.

\begin{figure}
	\begin{center}
		\vspace{-0.4in}
		\includegraphics [width=1.0\linewidth]{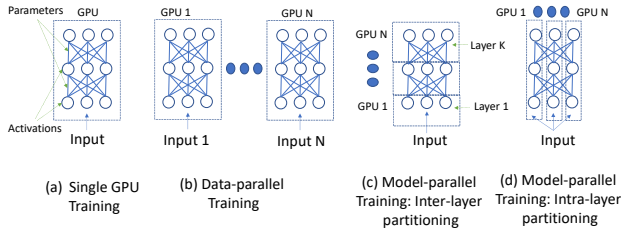}
		\vspace{-1.1in}
		\caption{{\bf Various architectures for training DLT jobs}}
		\label{fig:models}
		\vspace{-0.3in}
	\end{center}
\end{figure}

\noindent
{\bf Data-parallel Training.} A straightforward way to reduce training time is to replicate the model in N GPUs as shown in Figure~\ref{fig:models}(b). Each GPU performs forward and backward passes of each iteration independently but then use a technique called all-reduce~\cite{allreduce} to synchronously compute an average gradient, that is then used to update the model.

\noindent
{\bf Model-parallel Training. } While data-parallel is mainly used to speed up training, model-parallel training helps with fitting massive models by spreading the parameters of a single model across multiple GPUs. Model-parallel training can be inter-layer and/or intra-layer.

\noindent
{\bf Inter-layer or pipeline parallelism.} A deep learning model is divided into a number of layers. Thus, a natural way to split a model among multiple GPUs is to distribute different layers among them. This is known as inter-layer or pipeline partitioning and is shown in Figure~\ref{fig:models}(c).

Gpipe~\cite{Gpipe} was one of the first systems to use pipeline partitioning to split a large model (6 billion parameters) among a number of GPUs. Gpipe follows the {\it synchronous SGD} semantics where gradients are applied synchronously at the end of a mini-batch but this can result in "bubble overhead" where many pipeline stages are stalled. Gpipe uses micro-batches to reduce the pipeline overhead but assumes that the user provides an optimized, partitioned model as input.

Unlike Gpipe, PipeDream~\cite{pipedream} achieves 100\% pipeline utilization in steady state. To achieve this, PipeDream sacrifices synchronous SGD semantics as the parameter update is stale/delayed. However, the cost of stale updates is that some models may fail to converge. For example, authors in~\cite{pipemare} show that PipeDream fails to converge on certain language translation tasks. In addition, PipeDream keeps {\it P copies of parameters} where P is pipeline depth, which limits the size of the model that can be trained. PipeDream-2BW~\cite{narayanan2020memory} reduces the number of copies to 2 but still suffers from stale updates.

PipeMare~\cite{pipemare} tries to address the convergence problem with PipeDream by introducing new, approximate techniques to correct for stale updates.  While the authors show that PipeMare has better convergence than PipeDream, they evaluate only on smaller models and haven't released their code for comparison. PipeMare also uses 25-33\% more memory than Gpipe which is an issue for larger models.

DAPPLE~\cite{fan2021dapple} maintains {\it synchronous SGD} semantics and strictly interleaves forward and backward passes in its pipeline schedule. This avoids recomputing activations, but requires each stage to store intermediate activations for multiple micro-batches until the pipeline steady state is reached.  This is infeasible for large models that require long pipelines.  The largest model that DAPPLE shows performance speedup for is Bert-48 with 600M parameters.

\noindent
{\bf Intra-layer parallelism.}
An alternative model-parallel approach is intra-layer partitioning where a single layer of a model is split across multiple GPUs as shown in Figure~\ref{fig:models}(d)
Mesh-Tensorflow~\cite{meshtf}, Megatron~\cite{shoeybi2019megatron}, and Turing~\cite{Turing} adopt this approach (Megatron recently has added support for pipeline-parallelism~\cite{narayanan2021efficient}). Low network latency/jitter and high network bandwidth are critical requirements for this approach since a large matrix multiplication within a layer is now split among multiple GPUs. Thus, for efficiency, Megatron and Turing models use DGX-2s with the model partitioned only within the DGX-2 so that the communication can benefit from the 2.4 Tbps NVlink connectivity.

\noindent
{\bf Memory optimization.} Another aspect to fitting large models is optimizing GPU memory usage. A model with $N$ parameters will need up to $16*N$ bytes of memory to store parameters and optimizer state~\cite{zero}. In addition, one needs the intermediate outputs of the forward computation, called the activations. The activation size depends on the model and batch size used and can be significant. Activations can be stashed in CPU memory and brought back to GPU as needed~\cite{geeps} but this can incur substantial overhead~\cite{cpumemoverhead}. Instead, Gradient checkpointing~\cite{checkpointing, checkmate} is used, where the intermediate activations are not saved during forward pass and are recomputed (based on saved input activations for each layer) for the backward pass. Since forward pass takes about one-third of the iteration compute, this adds about 33\% overhead. This approach is used by most systems including Gpipe, Megatron and \varuna\ to train large models.

ZeRO/DeepSpeed~\cite{zero} optimizes memory usage in data-parallel training by sharding the redundant state among the replicas. This is complementary to systems like Gpipe and \varuna. In addition, Zero/DeepSpeed started with intra-layer and added inter-layer parallelism to fit massive models. Zero-infinity~\cite{zero-infinity} extends Zero to carefully offload GPU memory to CPU and SSD, thereby enabling scaling to models with trillions of parameters.

\noindent
{\bf Resource elasticity and Spot VMs.} Recent frameworks such as PyTorch Elastic~\cite{elastic} and MXNet Dyanmic~\cite{mxnetdl} support training over dynamic set of resources. However, these frameworks simply pass the burden of adapting to elasticity to the user who has to ensure that their scripts are configured with the right parameters for any resource availability. Authors in both~\cite{or2020resource} and ~\cite{harlap2017proteus} support auto-scaling but only for jobs that fit within a single-GPU. They increase the degree of data-parallelism (N) as long as the marginal utility of increase is higher than marginal cost. \varuna\ handles scaling automatically for massive jobs that use both model-parallel and data-parallel training.
Proteus\cite{harlap2017proteus} enables training models over a mixed set of spot and dedicated VMs.  But this approach is specific to parameter server framework and does not scale with large cluster and model sizes. Spotnik\cite{wagenlander2020spotnik} introduces an adaptive all-reduce method for spot instances, but only works for smaller image models and focuses on finding the best 2-D configuration that has the highest efficiency.

A summary of the various systems for training massive models is shown in Table~\ref{tab:arch}.  First, most systems initially started as either intra-layer or inter-layer but Megatron and Deepspeed have recently added inter-layer support (denoted by *), supporting both modes of model parallelism. Second, notice that only PipeDream and PipeMare do not follow synchronous SGD semantics and are thus susceptible to convergence issues. Third, systems that use intra-layer parallelism like  Mesh-Tensorflow handle the partitioning automatically for the user. On the other hand, with Gpipe/DeepSpeed, a significant burden of partitioning the model is placed on the user. PipeDream has support for automatic partitioning based on profiling. However, they only support a set of whitelisted functions while models like BERT use several custom functions. Thus, one would need to rewrite such models to work with PipeDream. \varuna\ automatically annotates a model with cut-points for re-configuration and adds a tracer that identifies potential shared variables that may need to be synchronized, easing programming for the user. Finally, only \varuna\ enables training over low-priority VMs through its elasticity support.

\begin{table}
	\small
	\begin{center}
		\begin{tabular}{| l | c | c | c|  c| c|  }
			\hline
			System  &  Intra- & Inter- & Sync- & User & Low- \\
			&  Layer  & Layer & SGD    & Ease & Pri. \\
			\hline
			Mesh-Tensorflow  & \checkmark & \text{\sffamily X} & \checkmark & \checkmark & \text{\sffamily X} \\\hline
			Megatron/Turing & \checkmark & \checkmark* & \checkmark & \checkmark & \text{\sffamily X}\\\hline
			Gpipe & \text{\sffamily X} & \checkmark & \checkmark & \text{\sffamily X} & \text{\sffamily X} \\\hline
			Pipe(Dream/Mare) & \text{\sffamily X} & \checkmark & \text{\sffamily X} & \checkmark* & \text{\sffamily X} \\\hline
			Zero/DeepSpeed & \checkmark & \checkmark* & \checkmark & \text{\sffamily X} & \text{\sffamily X} \\\hline
			\varuna & \text{\sffamily X} & \checkmark & \checkmark & \checkmark & \checkmark \\
			\hline
		\end{tabular}
	\end{center}
	\caption{{\bf Systems for training massive models: Features}}
	\vspace{-0.4in}
	\label{tab:arch}
\end{table}

\section{Design Overview}
\label{sec-design}

In this section, we describe the design of \varuna, highlighting how it handles the constraints of commodity networks.

\subsection{High-level architecture}

\begin{figure}
	\begin{center}
		\vspace{-0.4in}
		\includegraphics [width=0.8\linewidth]{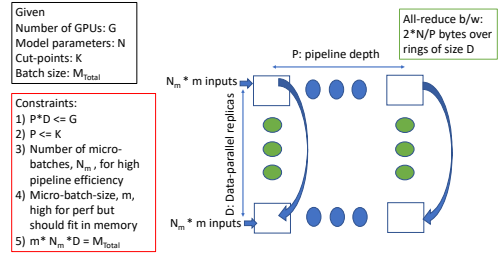}
		\vspace{-0.5in}
		\caption{{\bf Problem setting and constraints}}
		\label{fig:problem}
		\vspace{-0.4in}
	\end{center}
\end{figure}

\varuna\ uses a combination of data parallelism and {\em pipeline} model parallelism as shown in Figure~\ref{fig:problem}.  Each model with $N$ parameters is partitioned into $P$ partitions, and each partition has multiple replicas, running like a data-parallel job.   Across the different partitions of the model, \varuna\ uses {\em pipeline parallelism}, where each stage $k$ (other than the first and last stage) gets input activations from the previous stage $k-1$, performs the forward pass computation, and sends the output activations to the next stage $k+1$.   Similarly, stage $k$ receives gradients from stage $k+1$, performs the backward pass computation, and sends input gradients to stage $k-1$.  

The backward computation requires the activations computed in forward pass, in order to compute the gradients for the previous stage.  However, as described in \S~\ref{sec-background}, activations take up a large amount of memory and hence massive models cannot remember activations. Instead,  \varuna\ recomputes activations by re-running the forward computation~\cite{checkpointing} using the stored input activation for each layer. Input activation is a fraction of model size, e.g., for 2.5B GPT-2, this is only 3.75~MB per input example.

\noindent
{\bf Observation 1: Pipeline parallelism instead of intra-layer parallelism.} Conventional wisdom today is that intra-layer model-parallelism should be used to scale up to the number of GPUs within a server and only then pipeline parallelism be used (See takeaway 1 in~\cite{narayanan2021efficient}; Deepspeed and Megatron both started with only intra-layer parallelism). 

In intra-layer parallelism, the matrix-multiplication computation in each layer is split between multiple GPUs. This requires two allreduces each in the forward, backward, and recompute passes for each layer. For each such allreduce, every GPU transfers 2 x hiddenSize x sequenceLength 16-bit floats in mixed precision training. For GPT-2 2.5B model with 54 layers, a hiddenSize of 1920, and a sequenceLength of 1024, the amount of data transferred is 2.4~GB/example/GPU.
Further, these all-reduces are {\it synchronous}, which implies GPU computations wait until communication completes. In contrast, pipeline parallelism only communicates end of layer activations+gradients which are 7.5~MB/GPU/example for the same model ($\approx300\times$ smaller) and this communication can overlap with computation. Thus, \varuna\ eschews intra-layer parallelism in favor of pipeline parallelism. We show that \varuna\ outperforms intra-layer parallelism not only in low-priority settings but even in hypercluster settings.

\noindent
{\bf Observation 2: Balancing pipeline overhead and all-reduce bandwidth.} Pipeline bubble overhead is directly proportional to number of pipeline stages (P)~\cite{Gpipe}~\cite{fan2021dapple}. Thus, conventional wisdom is to minimize the number of pipeline stages (also why intra-layer is preferred today). However, when P is reduced, the all-reduce bandwidth (2N/P) increases. Further, D = G/P also increases, resulting in significantly higher network load. While this bandwidth increase is typically not an issue in hypercluster settings, it can significantly impact performance in commodity networks. Therefore, \varuna\ identifies an optimal balance between P and D that is unique to the low priority setting (Section~\ref{sec-lowpri}).

\noindent
{\bf Observation 3: Tolerating higher latency/jitter.} Unlike hypercluster, commodity networks can suffer from high latency/jitter. \varuna\ addresses this constraint in several ways. First, \varuna\ uses a large number of {\em micro-batches}~\cite{Gpipe} to overlap communication with computation, thus moving latency off the critical path. Second, \varuna's pipeline schedule (Section~\ref{subsec:pipeline_efficiency}) is specifically designed to be able to opportunistically adapt to network jitter. Finally, \varuna\ explicitly profiles latency and jitter, and incorporates it in its simulation (Section~\ref{sec-lowpri}) to identify the best parallelism configuration.

\begin{figure}
	\begin{center}
		\vspace{-0.2in}
		\includegraphics [width=0.8\linewidth]{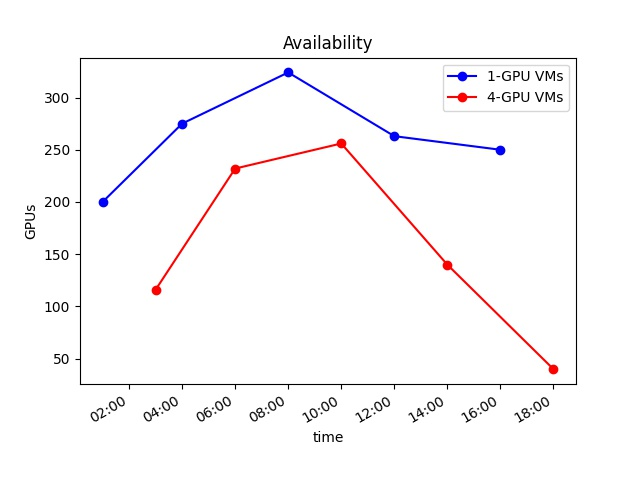}
		\vspace{-24pt}
		\caption{{\bf Availability of 1 and 4 GPU VMs}}
		\label{fig:1vs4gpu_availability}
		\vspace{-0.2in}
	\end{center}
\end{figure}

\noindent
{\bf Observation 4: Single GPU vs Multi-GPU VMs}. Our experiments show that for {\em low-priority VMs}, single GPU VMs are more readily available than 4-GPU VMs. For example, Figure~\ref{fig:1vs4gpu_availability} shows aggregate GPU availability when low-priority VMs with 1 and 4 GPUs are requested/released alternately in Azure over a duration of 16 hours. Thus, systems that can work with 1-GPU VMs can utilize higher aggregate capacity, albeit at the cost of stressing the network even more (PCIe replaced by Ethernet). Due to \varuna's thrifty use of networking resources, \varuna\ is able to train on 1-GPU VMs at almost the same performance (2\% difference) as 4-GPU VMs, thus enabling faster training completion.

\subsection{Tuning pipeline efficiency}
\label{subsec:pipeline_efficiency}
As in any pipeline, the efficiency of \varuna\ depends on reducing pipeline stalls.  Pipedream~\cite{pipedream} reduced stalls by sacrificing the semantics of synchronous stochastic gradient descent (SGD).  \varuna\ follows the GPipe~\cite{Gpipe} approach that preserves the semantics of sync-SGD but uses a {\it novel schedule that is more efficient and more tolerant of network jitter}.

\varuna\ uses a combination of a static rule-based schedule enumerated for a given pipeline depth, along with an opportunistic policy that is employed to hide jitter.   The rule-based schedule is generated based on a tool that enforces the following constraints:

\begin{enumerate}[nosep]
\item
If stage $k$ will complete its backward pass for micro-batch $m$ in time $t$, the recompute of gradients must be scheduled in stage $k-1$ at time $t^{\prime}$ such that $ t - t^{\prime} > T_f $, where $T_f$ is the time taken per forward pass per stage (assuming uniform stages)

\item
If a recompute has completed for stage $k$ for micro-batch $m$, unconditionally wait for the corresponding backward task for $m$ to be scheduled (as running a forward pass will create another set of activations, taking 2x the memory for activations)

\item
If a stage $k$ has both forward and backward tasks ready to be scheduled, prefer the backward task.

\end{enumerate}

Each stage sticks to the above pre-defined offline schedule.   Sometimes, the schedule for stage $k$ may indicate that the backward pass for micro-batch $m$ must be scheduled, but the gradients for $m$ may not have arrived yet from stage $k+1$; in those cases, \varuna\ deviates from the schedule and opportunistically schedules another ready task (\eg, forward pass for a micro-batch).  This allows \varuna\ to be work-conserving during network jitter, reducing pipeline stalls.

\begin{figure}[t]
	\center{\includegraphics[width=3.3in]
		{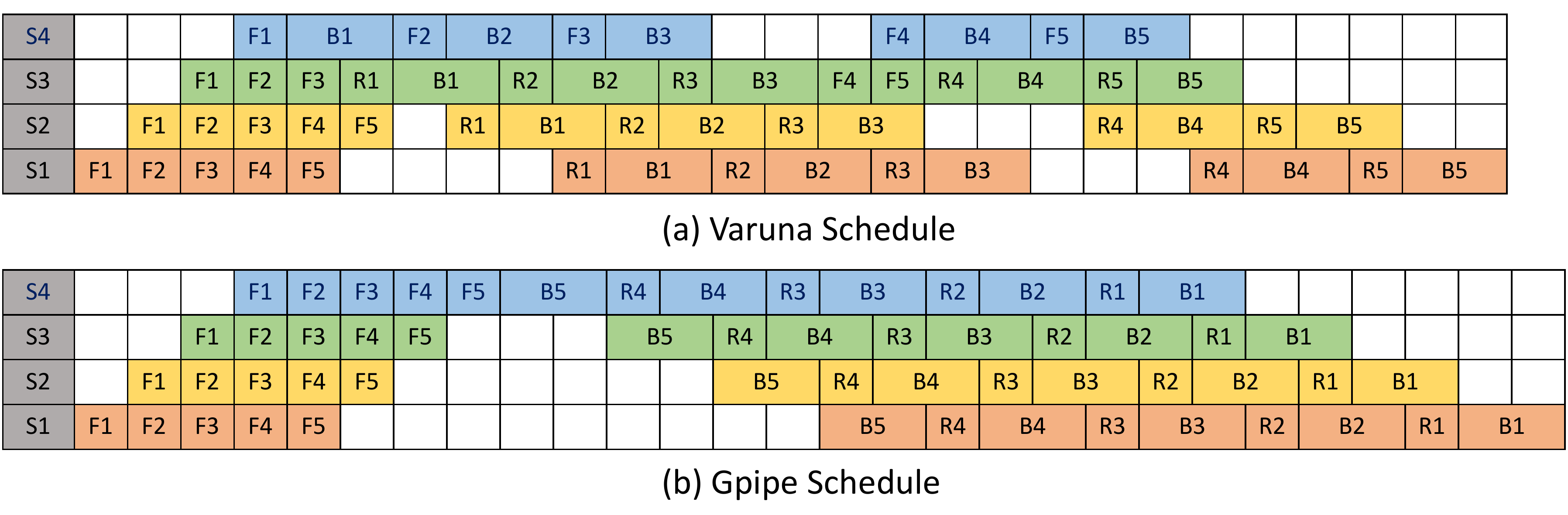}}
	\vspace{-12pt}
	\caption{\label{fig:varunavsgpipe} {\bf \varuna's micro-batch schedule contrasted against Gpipe. S denotes the stage, F denotes forward, B denotes backward and takes twice as long as forward, and R denotes recompute, with number representing micro-batches (1 to 5)}}
	\vspace{-0.25in}
\end{figure}

A qualitative comparison of \varuna's {\em static} schedule against GPipe's schedule for a 4-stage pipeline with 5 micro-batches is shown in Figure~\ref{fig:varunavsgpipe}. First, note that  \varuna\ is more efficient overall and uses 1 less time unit compared to Gpipe as it has fewer stalls (white space). Second, the whitespace in \varuna\ is distributed throughout the schedule while it is concentrated in the middle in Gpipe. This makes Gpipe's schedule more vulnerable to network jitter as there is no free time slots later in the schedule that can serve as a buffer. Third, the last stage (S4) in \varuna\ does not perform any recompute unlike S4 in Gpipe (which only avoids recompute for the fifth micro-batch). Language models have final embedding layers that are less compute intensive than the rest of the layers. Avoiding last-stage recompute allows \varuna\ to pack in such embedding layers in the final stage itself, improving efficiency without upsetting the pipeline balance. Finally, note that forward passes are interspersed in \varuna\ throughout the schedule (see stage 3). This allows \varuna\ to make use of opportunistic scheduling to handle jitter, unlike Gpipe where all the forwards are bunched in the beginning. 

Pipeline parallelism has been recently introduced into Deepspeed~\cite{zero} and Megatron~\cite{narayanan2021efficient}, with their variations of the Gpipe schedule.  As we show in Section~\ref{sec-evaluate}, \varuna's schedule is more efficient than Gpipe, Deepspeed and Megatron's pipeline schedules in commodity network settings.

\section {Handling pre-emptions: Job Morphing}
\label{sec-lowpri}

An important goal of \varuna\ is to reduce cost by harnessing {\em low-priority VMs} that cost 4-5x cheaper than dedicated GPU VMs.  As spot VMs can be pre-empted often, \varuna\ needs to adapt to a variable amount of resources. \varuna\ uses a novel technique of {\em job morphing} to dynamically configure the job to run at best performance with available resources.

Previous approaches for elasticity in DLT jobs mostly address data-parallel jobs~\cite{elastic,mxnetdl,or2020resource}.   Pipedream~\cite{pipedream} performs automated placement, but relies on profiling ~1000 mini-batches, and an expensive optimizer whose cost is O($N^2*L^3$) where N is the number of GPUs and L is the number of layers.   Such cost is clearly prohibitive for large models.  Even for much smaller models and at most 16 GPUs, their optimizer takes ~8s and ``a few minutes" for the profiling.  \varuna\ scales to models that are 100x larger and utilizes hundreds/thousands of GPUs, thus, a single run of pipedream's optimizer will take several hours.   More recent work~\cite{narayanan2020memory} uses end-to-end profiling for each configuration of the job; the cost of such profiling can perhaps be tolerated in their setting where the number of GPUs is fixed, but in the context of low-pri VMs that \varuna\ targets, frequent auto-configs need to be performed on preemptions (not just once at job startup). 

In contrast, \varuna\ uses a novel approach of {\em scale-invariant calibration} and {\em parametrized simulation} to identify the best configuration.  Our approach ensures that the profiling state space is kept to a minimum and importantly, is done only once at startup, rather than being repeated every time the number of GPUs change due to pre-emptions.

\subsection{Problem setting and constraints}

Figure~\ref{fig:problem} depicts the problem setting and the main constraints for \varuna.
At any given time, \varuna\ identifies $G$ GPUs that are available as spot instances. It then
needs to identify configuration parameters, pipeline depth $P$ and data-parallel replicas $D$ such
that $P*D \le G$. Further, $P$ has to be less than $K$, the number of cut-points specified by the user.
The next constraint is selecting $N_m$, the number of micro-batches for pipeline efficiency.  Then, one needs to select $m$, the micro-batch size for best performance while respecting the total batch size constraint $M_{Total}$ supplied by the user.

For a per-GPU micro-batch size of $m$, and $G$ GPUs, the total mini-batch size $M_{Total}$ is $m * N_m * G$.   The value of $m$ is constrained at both ends: $m$ cannot be too small, as it  reduces efficiency of the CUDA/tensor operations within the GPU (in BERT-large~\cite{bert-large}, $m = 8$  performs 26\% better than $m = 4$); 	$m$ cannot also be too large, as it then cannot fit into the GPU memory of a single GPU. 

\subsection{Correctness-preserving morphing}

Unlike frameworks that require the user to specify a different set of hyper-parameters for each configuration~\cite{elastic,mxnetdl}, \varuna\ keeps $M_{Total}$ fixed across configurations, thus easing user burden.   The user simply writes the script for a mini-batch size $M_{Total}$ that ensures {\em maximum parallelism}, and \varuna\  apportions it among dynamically varying resources.

However, this places pressure on the choice of $m$; if $M_{Total}$ was calibrated for a large number of GPUs, one cannot fit it in fewer GPUs as it would increase $m$ beyond the memory of single GPU.  Previous work~\cite{or2020resource} deals with this problem by setting a conservative small value for $M_{total}$ resulting in poor performance at large scale (because $m$ gets too small).  \varuna\ performs automatic gradient accumulation {\em within} a GPU.   When resources shrink resulting in a high value of $m$ that cannot fit in memory, \varuna\ simply increases $N_m$, the number of micro-batches.

\subsection{Scale-invariant Calibration}

Morphing in \varuna\ uses a one-time profiling step to calibrate primitive parameters of the hardware and the model.  Key to making this scalable is that the primitive parameters are chosen to be (a) mutually orthogonal (so the parameters can be calibrated in parallel, reducing latency); (b) agnostic to the end-to-end configuration of the model (so the number of parameters is small); (c) independent of the total number of GPUs (so that it is scale-invariant when $G$ changes).  This makes the calibration task independent of the size of the configuration state-space, in contrast to end-to-end profiling~\cite{narayanan2020memory,pipedream} where the state space is much larger.

The parameters that are calibrated in this phase are listed in Table~\ref{tab-parameters}.  Scale-invariant calibration exploits the mini-batch predictability of DLT job execution~\cite{xiao2018gandiva, sivathanu2019astra} to measure these just once.  Note that $F_i$ and $B_i$ are independent in the $i$ and $m$ dimension, and are measured in parallel on multiple GPUs by running a few micro-batches using random input values to mock the previous stages. The network times are measured for sending each activation size, both intra- and cross-node.
Similarly, $AR_i$  is also measured independently by using a profiling allreduce run that uses the same number of gradients as that of a cut-point for different ring sizes.  To model the scenario where multiple stages of the pipeline will be in the same node (and hence all stages could be performing allreduce in parallel), our micro-benchmark measures the allreduce when there are $k$ allreduces in flight (where $k$ is the number of GPUs per node).   
The time taken for collecting all these measurements is simply the time for a few micro-batches, and is under a minute for even a 10 billion parameter model.  Further, it is independent of the total number of GPUs that the job may be scheduled in.

\begin{table}
	\small
	\begin{center}
		\begin{tabular}{| l | l | }
			\hline
			Parameter &  Description  \\
			\hline
			$F_i(m)$ & Forward-pass compute-time for $C_i$ \\
         $B_i(m) $ & Backward-pass compute-time for $C_i$  \\
         $Act_{intra}^i (m)$ & Latency (same node) to send activations of $C_i$ \\
         $Grad_{intra}^i (m)$ & Latency (same node) to send gradients of $C_i$ \\
         $Act_{inter}^i$ (m) & Latency (cross-node) to send activations of $C_i$  \\
         $Grad_{inter}^i$ (m) & Latency (cross-node) to send gradients of $C_i$ \\
		   $ AR_i (D) $ & Gradient Allreduce time for $C_i$ on ring size D  \\
			\hline
		\end{tabular}
	\end{center}
	\caption{{\bf Primitive parameters for calibration.  $C_i$ represents the $i$th cut-point. $m$ is the micro-batch size. All network times except $AR_i$ include mean latency and jitter.}}
	\label{tab-parameters}
	\vspace{-0.2in}
\end{table}

\subsection{Parametrized Simulation}

Once the primitive parameters are calibrated, they are then fed into an event-driven simulator that models the execution of \varuna.  Unlike calibration, the simulator needs to run for each configuration:  the simulator takes in values for $G$, $P$, and $D$, and also the mapping of cut-points to stages (in the case of homogeneous-block models such as GPT-2, the mapping is uniform across stages), and simulates one full mini-batch ($N_m$ micro-batches followed by the allreduce), and outputs the estimated time-per-minibatch.  As we show in Section~\ref{sec-evaluate}, the simulated times are within 5\% of the actual measured times for that configuration, showing the sufficiency of the parameters for calibration.

To reduce the state space of options to feed to the simulator, \varuna\ first picks
the lowest $m$ at which $F_i(m)/m$ stops improving. It then sweeps through all
$P$ starting with the smallest $P$ where the model fits, and increase it until the maximum number of cut-points or $G$. For each $P$, it picks only one assignment of cutpoints to stages such that they are balanced in $F_i(m)$. Thus, the total exploration size is at O(G).  Finally, note that identifying $m$ needs to be done only once as this can be reused in subsequent morphing decisions.

An interesting result from the auto-morphing (validated with real runs) is that contrary to popular belief, a shallow pipeline does not always perform better than a deep pipeline (Observation 2 in \S~\ref{sec-design}).  While smaller number of pipeline stages (P) helps with overlapping inter-stage network latency with intra-stage compute, there is a tradeoff; as $G = D x P$, for a given $G$, a small $P$ results in a larger $D$; a large $D$ incurs a high cost for performing the data-parallel allreduce communication.  Thus, in cases of large $G$, a deeper pipeline (larger $P$) keeps $D$ small.  Table~\ref{table-pipeline-depth} shows performance from a real run of training a 2.5 billion parameter model with 6-way and 9-way pipelining; the optimal pipeline depth varies with G.  The parametrized simulation helps us detect such cases and prefer the best performing configuration for a given G.
Another side effect of variable number of GPUs is that few GPUs may be left unused; for example, with 100 total GPUs, 6-way pipelining can only use 96 GPUs while 9-way pipelining can use 99; thus, the difference in total throughput is larger than the difference in normalized per-GPU throughput.

\begin{table}
	\small
	\begin{center}
		\begin{tabular}{| l | c | c | c | }
			\hline
			Num GPUs  &  Config (PxD) & Total Ex/s & Ex/s/GPU  \\
			\hline
			36 & 6x6   & 66.60  & 1.85 \\\hline
			36 & 9x4   & 65.88  & 1.83 \\\hline          
			36 & 18x2  & 50.04  & 1.39 \\\hline
			100 & 6x16 & 155.52 & 1.62 \\\hline
			100 & 9x11 & 164.34 & 1.66 \\\hline
			100 & 18x5 & 99.00  & 1.1 \\\hline
		\end{tabular}
	\end{center}
	\caption{{\bf Sensitivity to pipeline depth (P) training a 2.5B GPT2 model. Ex/s is throughput (examples/sec)}}
	\label{table-pipeline-depth}
	\vspace{-0.2in}
\end{table}

\subsection{Continuous Checkpointing}

To handle unexpected pre-emptions, \varuna\ constantly checkpoints the model state across all stages. Each layer is checkpointed independently. Since data-parallel replicas have the same model state, we shard the checkpointing across replicas for performance. For consistency across the pipeline stages, the checkpointing is done at the end of a mini-batch, every few mini-batches.  When the configuration changes, \varuna\ can resume the job from the latest checkpoint, with the new configuration; as each layer is checkpointed separately, it allows the morphing framework to even use a different mapping of layers to stages (\eg, if the pipeline depth has to reduce). To reduce impact on training time, the checkpoints are written to local SSD of the VM, and copied to cloud storage in the background.

\subsection{\varuna\ Manager}
\label{subsec:varunamanager}

The \varuna\ manager runs on a dedicated VM, and monitors the tasks running on different GPUs.  The manager also decides on the placement of the stages and replicas of a job.

\noindent {\bf Handling fail-stutter machines: }
When running on pre-emptible VMs, we repeatedly encountered instances where a particular VM or GPU would perform slower than the rest (often by as much 30\%).  Because of the synchronous nature of DLT jobs, even a single slow GPU would slow down the entire job.  Fortunately, as data parallel replicas of the same stage run the same computation, such {\em fail-stutter} behavior is easy to correct.  Each task sends a {\em heartbeat} to the manager that contains the GPU compute time per micro-batch for the forward and backward pass.  If the manager detects any outliers, it omits that VM when scheduling task replicas.

\noindent {\bf Tracking size of cluster: }
As spot VMs can get pre-empted at any time, the manager also detects preemptions when it has not received a heartbeat from a VM, and triggers the morphing functionality to reconfigure the job. Similarly, the manager periodically keeps trying to grow the cluster by invoking the appropriate provisioning APIs in the cloud.

\section{Enabling Ease of Programming}
\label{sec-prog-ease}

A key challenge in model parallelism is making it easy to use for ML developers.  Existing approaches like GPipe~\cite{Gpipe} and DeepSpeed~\cite{zero} require the user to rewrite their model using specific libraries or in cumbersome ways (\eg, flatten the model into a linear set of layers).   Ideally, the framework should not require any changes to the model, and allow the programmer to write the model as if it runs on a single GPU.    PipeDream~\cite{pipedream} attempts to perform such automatic model splitting, but is tied to specific libraries.   PipeDream re-implements a whitelisted set of pytorch operations and make use of {\tt TensorWrapper}s to get a sequential list of all computations.  This is not scalable to large models like BERT with a high number of custom Python functions, branching and 1000s of lines of code; a standard implementation of BERT crashes on PipeDream. PipeDream-2BW's publicly available code is implemented by heavily customizing the Megatron repository.

It is therefore not surprising that intra-layer parallelism is more popular because the user does not have to worry about the partitioning.  For example, while Mesh-Tensorflow~\cite{meshtf} is actively maintained and supported by Google~\cite{meshtf-source,meshtf-talk}, the GPipe implementation~\cite{gpipe-source} is quite primitive and works only within a single node.  Similarly, Nvidia's Megatron~\cite{shoeybi2019megatron} and Microsoft's DeepSpeed~\cite{zero} all started with intra-layer partitioning and only recently have added pipeline support.

\subsection{Auto-partitioning}

A key observation behind \varuna's auto-partitioning approach is that massive models such as BERT-large, GPT-2 or ResNet-150 {\it inherently use repetitive structures}; the same block of layers is repeated multiple times to scale models.  For example, in BERT-large or GPT-2, a {\em transformer} encoder/decoder is the basic building block~\cite{Transformers} (referred to as layer in this paper), which is repeated 24 and 48 times, respectively. If one can identify low-activation sizes within each block, they can serve as {\it cut-points} where the model can be partitioned.

\varuna\ addresses ease-of-use by automatically partitioning models in two steps: identifying suitable cut-points via model profiling, and then activating a subset of these at run time based on resource availability. Cut-points are "cuts" in the model that slice the computation into equally heavy code sections ending with low activation sizes.  These fine-grained sections can be combined at run time for various pipeline depths up to the total number of cut-points.

Cut-points are identified by profiling the model for execution times and activation sizes for each operation. \varuna\ also checks that there is no overlap of parameters across cut-point boundaries, and parameters that are reused across boundaries are marked as shared parameters (e.g., embedding weights in transformer models). Based on the desired number of cut-points, \varuna\ uses compute time to shortlist end points for each code section, and picks those with lowest activation size to maintain a high compute-communication ratio.   At run time, based on the number of GPUs and bandwidth available, the optimal pipeline depth $P$ is estimated as described earlier and one or more cut-points are grouped together into $P$ partitions.

\subsection{Tracking cross-partition dependencies}

Another significant usability challenge with inter-layer partitioning is implicit data dependencies that span across partitions, but are important to preserve for model convergence and accuracy. Within the model these might be in the form of shared weights. For example, in the GPT-2 and BERT models, the embedding weights for the first and last layer are ``tied'', \ie, they are meant to use the same parameters. These are flagged by \varuna\ during automatic cut-point detection and synchronized during training. A more tricky scenario occurs when these dependencies are not in the user-written model code, but hidden in some third-party libraries the model uses. For example, the APEX library~\cite{APEX} for fp16 training performs loss scaling when it detects computation overflow in any layer. In a partitioned world, one stage may hit overflow while others may not, thus requiring an allreduce to synchronize it. Another example arises in optimizers such as NVLamb~\cite{NVLAMB} that use a ``global norm'' value computed across layers. The model writer may not even be aware of such sharing, so it is easy to miss them while partitioning, resulting in lower accuracy.

To address this problem, \varuna\ provides a tracer that detects such implicit data dependencies. The tracer performs a dry run of training, where model partitions are executed in the same process sequentially.  We make minor modifications to the PyTorch library to support a {\em profiling mode}, during which each created {\tt Tensor} is marked with a cut-point number to which it belongs.   In this mode, all python function calls are traced, and any function that uses tensors from more than one partition is flagged. Any tensors that are unmarked during the run are also considered "common" as they are created outside the model/Varuna scope.  These are then provided as a list of potential violations to the user, who can mark these as ``shared" in \varuna.  For all shared tensors, \varuna\ performs an allreduce that synchronizes them every mini-batch.   In the models we trained for convergence, the tracer caught all instances of such sharing.

\section{Implementation}
\label{sec-implementation}

In this section, we describe the implementation of some of the unique aspects of \varuna. Our current implementation is built on the PyTorch framework~\cite{pytorch}.

\noindent
{\bf Cut-points.} 
Based on the details gathered in the dry-run of \varuna\ at initialization time, the model is automatically partitioned according to its rank and total number of nodes in the pipeline. For example, if \varuna\ decides to partition BERT-Large into 4 GPUs, four equally spaced cut-points are activated in the model and the rest of the cut-points become pass through. For rank 0's forward pass, the initial set of modules process the input and when the first activated cut-point is reached, it is configured to send the output activation tensors to the rank 1 process. In rank 1, the first module in the forward pass is set to the cut-point that is awaiting tensors of pre-identified shape from rank 0. It then executes the modules following this until it reaches the next cut-point which is configured to send the activation tensors to rank 2 and so on. The backward pass executes in the reverse manner and sends the gradients tensors to the previous rank. In this way, \varuna\ takes in a user model annotated with cut-points and automatically partitions it into N GPUs to operate in a pipelined manner. If there are multiple data-parallel stages, those pipelines are also setup similarly.

\noindent
{\bf Overlapping computation and communication.}
The activation and gradient communication is in the critical path of the computation. \varuna\ overlaps this communication with computation so that communication overhead is minimized. Each rank spawns separate threads for sending and receiving activation and gradients. A queue interface is established between the cut-points and the sending/receiving thread. Each thread uses PyTorch's asynchronous sends and receives to transmit/receive tensors from/to the queue and the cut-points and modules independently compute on the tensors whenever they are available.

\noindent
{\bf Synchronization of shared state.}
\varuna\ establishes two process groups (or all-reduce rings) for each process. One process group consists of the data-parallel replicas of the same pipeline stage, over which an all-reduce is performed to get the average gradient. A second process group is established among all the processes of each pipeline. This is needed for synchronizing on data shared across partitions (\eg, flagged by the automatic debugger - \S~\ref{sec-prog-ease})

There are various other functions as well such as saving and restoring GPU random number state to ensure recomputation correctness, generating and utilizing the pipeline schedule, periodic checkpointing, resuming from checkpoints upon preemption, communicating forward/backward times for fail-stutter fault-tolerance, looking for new VMs to grow the cluster, \etc\ In total, \varuna\ consists of about 2400 lines of Python code and 550 lines of C++ code.

\section{Evaluation}
\label{sec-evaluate}

In this section, we first compare the performance and cost savings of \varuna\ against prior systems. Second, we highlight how \varuna\ is uniquely able to navigate the dynamism of spot VM availability while maintaining high training performance. Finally, we show how the performance gains of \varuna\ directly translates to faster time-to-convergence.

\noindent
{\bf Experimental setup.} For our experiments, we use two setups. Unless specified otherwise, experiments use a cluster of up to 300 GPUs, using low-priority spot VMs of type NC24\_v3 (4-GPU) or NC6\_v3 (1-GPU) in Azure.  Each 1-GPU VM has Nvidia Volta-100 GPU with 16GB memory, 6 Xeon cores, 112GB of CPU RAM and 10~Gbps ethernet. All VMs are allocated in a single region (South Central US), but have no other locality; in other words, the pair-wise connectivity between the VMs can be routed through multiple levels of bottleneck switches, that in practice limit bandwidth.  The second setup we use, that we refer to as ``hypercluster'', comprises of 16 Nvidia DGX-2 nodes, where each node has 16 V100 GPUs connected via NVLink. The 16 DGX-2 nodes are connected via 200 Gbps Infiniband.

For our workloads, we run the two most popular model architectures: BERT-large~\cite{bert76min}, and GPT-2~\cite{gpt-2}.  For BERT-large, we run a model with 340 million parameters, while for GPT-2, we run three configurations: a 2.5 billion parameter model, and a model with 8.3 billion parameters, both from Nvidia~\cite{shoeybi2019megatron}, and a 20 billion parameter model.  To  demonstrate scaling, we also show results from a massive model with 200 billion parameters. Note that the 170 billion parameter GPT-3 model is based on the same architecture as GPT-2. Finally, when comparing performance, we use the same mini-batch size for \varuna\ and other systems.

Note that although our evaluations are with language models, \varuna\ does not make any assumptions about the DNN, and will work for all models.  The workloads were picked solely due to their large sizes and prevalence in hyper-clusters today.

\subsection{Performance \& Cost}

We first evaluate \varuna\ on training performance. The baseline configuration is chosen based on the best reported configuration for each of the models: fully data-parallel training for BERT-Large, and data-parallel with intra-layer partitioning (Megatron~\cite{shoeybi2019megatron,Turing}) for GPT-2 models that cannot fit within the 16GB RAM of one GPU. We also compare with other pipeline architectures ~\cite{Gpipe,zero,pipedream,narayanan2021efficient}.

We report two metrics of performance for each of these experiments: the number of input examples that were processed per second per-GPU, and the per-GPU teraflops/sec. For the latter, we remove the 33\% cost of recompute so that only useful work is captured.

For \varuna\ and other pipeline schemes, we denote the configuration in the format $P\times D$ where $P$ is the number of pipeline stages, and $D$ is the degree of data parallelism per stage. As each model has a minimum value for $P$ (depending on how many layers can fit in a single GPU memory), the total number of GPUs used will be closest multiple of this $P$; e.g., if $P = 15$, \varuna\ would use only 60 GPUs out of 64.

\subsubsection{Comparison to Intra-Layer Partitioning}

We first compare \varuna\ with the intra-layer partitioning scheme of Megatron~\cite{shoeybi2019megatron,Turing}. We compare four configurations: (a) Megatron on commodity 4-GPU VMs (b) Megatron on hypercluster (c) \varuna\ on commodity 4-GPU VMs, and (d) \varuna\ on hypercluster. We also compare against fully data-parallel configuration for BERT-large model which can fit in a single GPU.

\begin{figure}
\centering{
\includegraphics[width=0.9\linewidth]{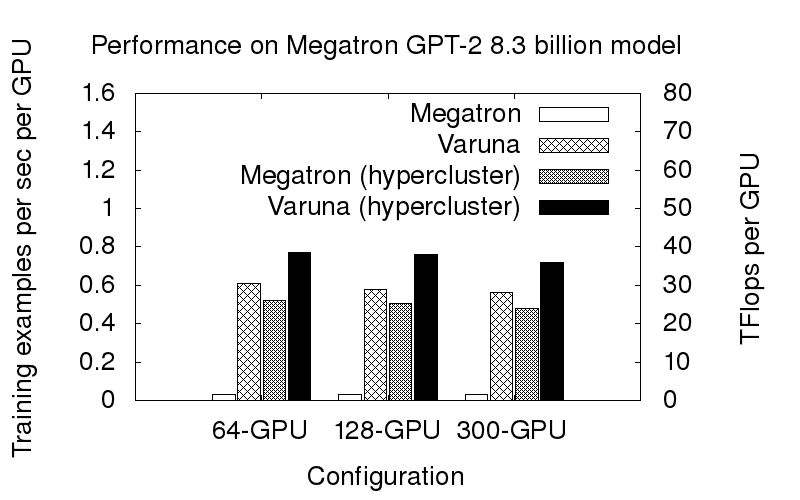}
\caption{{\bf Performance of \varuna\ and Megatron on the GPT-2 8.3 billion parameter model.}} 
\label{fig:megatron-8B}
\vspace{-0.2in}
}
\end{figure}

\noindent
{\bf GPT-2 8.3 billion model.}
Figure~\ref{fig:megatron-8B} compares the performance of \varuna\ with the Megatron baseline, both using a mini-batch size of 8192, under several configurations of the GPT-2 8.3 billion model - on 64 GPUs, 128 GPUs, and 300 GPUs. The corresponding number of GPUs used for \varuna\ were 54 (18x3), 126 (18x7), and 288 (18x16) respectively.  As can be seen, on commodity low-pri VMs the training speed (examples per second per GPU) with {\it \varuna\ is about 18x better than the Megatron on the same VMs}.

Figure~\ref{fig:megatron-8B} also shows the performance of Megatron and \varuna\ on a hyper-cluster environment.   Interestingly, \varuna\ (0.56 ex/s/GPU) on spot VMs performs 17\% {\em better} than Megatron (0.48) on hypercluster, {\em despite} running on commodity VMs which are 5x cheaper.  {\it The cost-performance is thus 5.85x better for \varuna}.  This demonstrates an inherent inefficiency with intra-layer partitioning even when using a high-speed network like NVLink! The reason why intra-layer partitioning performs worse is because of the large, synchronous allreduces during the forward/backward passes  (Observation 1). \varuna\ partitioning on the other hand overlaps GPU compute with communication between stages. The only idle time comes due to pipeline bubbles which are bounded by using a sufficient number of micro-batches and via our schedule. Not surprisingly, {\it \varuna\ on hypercluster performs even better (48\% faster) compared to Megatron}. Finally, prior work on Megatron (Table 2 in~\cite{shoeybi2019megatron}) used a smaller mini-batch size (512) and quotes a time of 2.1 days for 68,500 iterations on a similar cluster of 512 GPUs. This results in only 0.378 examples/s/GPU, lower than the 0.48 we report, as larger batch sizes are more efficient even for intra-layer parallelism.

\begin{figure}
\centering{
\includegraphics[width=0.9\linewidth]{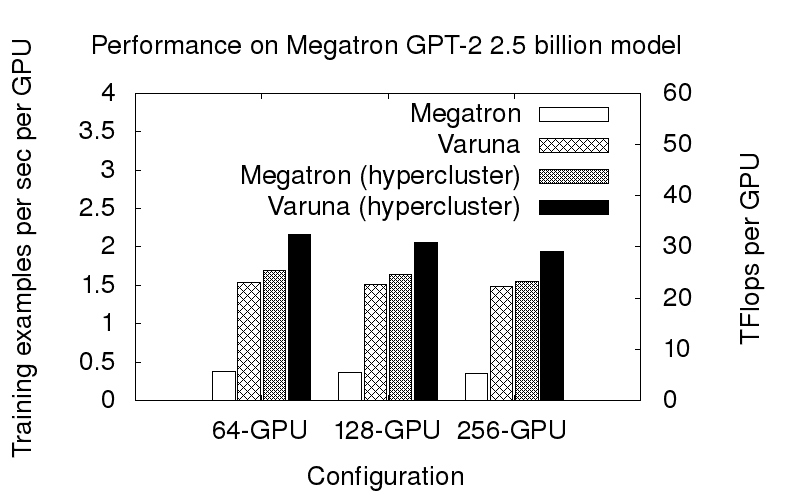}
\caption{{\bf Performance of \varuna\ and Megatron on the GPT-2 2.5 billion parameter model.}} 
\label{fig:megatron-2-5B}
\vspace{-0.15in}
}
\end{figure}

\noindent
{\bf GPT-2 2.5 billion model.}
Figure~\ref{fig:megatron-2-5B} compares \varuna\ with Megatron baseline for a smaller GPT-2 model with 2.5 billion parameters.  Again, \varuna\ performs 4.1x better than Megatron on commodity VMs, and gets within  4\% of hypercluster performance, for a performance-cost advantage of 4.8x. Further, \varuna\ on hypercluster performs better (25\% faster) compared to Megatron. The configs we use for \varuna\ are 9x7 (63 GPUs), 9x14 (126 GPUs) and 9x28 (252 GPUs) respectively.

\begin{figure}
\centering{
\includegraphics[width=0.8\linewidth]{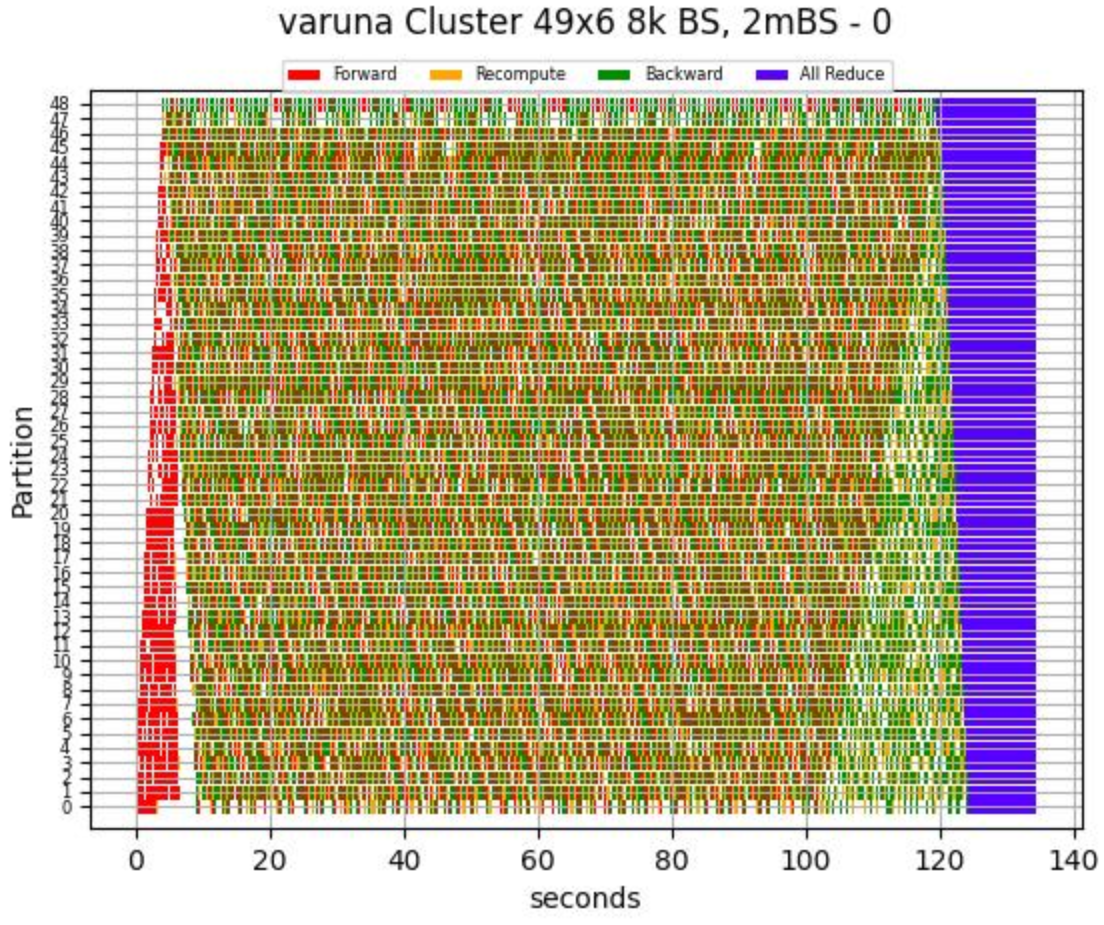}
\vspace{-0.1in}
\caption{{\bf Execution of \varuna\ mini-batch on the GPT-2 20B model (49x6 config).}} 
\label{fig:gantt-20b}
\vspace{-0.1in}
}
\end{figure}

\begin{table}
	\small
	\begin{center}
		\begin{tabular}{| l | c | c | c | }
			\hline
			System  &  Num  & Ex/s & TFlops/s \\
			  &  GPUs & /GPU & /GPU \\
			\hline
			20B  \varuna\ (LP) & 294  & 0.2 & 25 \\\hline
			19.2B Megatron (HC) & 256 & 0.112 & 14 \\\hline
			20B Megatron (HC) & 256 & 0.015 & 1.9 \\\hline
			20B Varuna (HC)  & 256 & 0.257 & 32.1  \\\hline
			\hline
		\end{tabular}
	\end{center}
	\caption{{\bf Comparison between \varuna\ and Megatron for 20B model.  LP is low-priority VMs, and HC is hypercluster. All configs use a mini-batch size of 8192.}}
	\label{table-20b}
	\vspace{-0.2in}
\end{table}

\noindent
{\bf 20 billion and 200 billion parameter models.}
To illustrate the ability of \varuna\ to scale to much larger models, we show in Table~\ref{table-20b}, the performance of \varuna\ on a GPT-2 model with 20 billion parameters (96 layers). For \varuna\ on low-pri GPUs, we used a 49x6 configuration to train this model on 294 GPUs. For comparison, Megatron on hypercluster could fit only a 19.2 billion parameter model with 16-way model parallelism (so that the intra-layer partitions stay within a single DGX-2 node); but even in this case, \varuna\ on commodity VMs performs 78\% faster than Megatron on hypercluster.  When we forced Megatron to 20B by 18-way model partitioning, its performance dropped by 10x.   Again, \varuna\ on hypercluster shows even better performance than \varuna\ on low-pri VMs.  

For more insight into the execution of \varuna, we show a detailed timeline of execution of a mini-batch in 49x6 configuration across different stages of the pipeline, in the form of a Gantt Chart~\cite{gantt}.  The chart in Figure~\ref{fig:gantt-20b} plots one of the 6 replicas.  Each horizontal line in the graph corresponds to a stage of the pipeline.  The purple region at the far right indicates the time for stage-wise 6-way allreduce.  Red bars correspond to forward pass, green bars pertain to backward pass, and orange bars indicate recompute.

To demonstrate extreme scale, we also ran a 200 billion parameter model on \varuna, with 100 layers and hidden size of 12960 on 102 GPUs with 102 pipeline stages and no data parallelism. Because of the large layer size, we run this with a micro-batch size of 1 and total batch-size of 512. For this model, \varuna\ keeps the optimizer state in CPU and performs GPU-CPU transfers at the end of mini-batch; the numbers reported include this cost. This ran at 0.022 ex/s/GPU, or 27.3 TFlops/s/GPU.

\noindent
{\bf BERT-large 340 million model.}
BERT-large is the smallest model we evaluate in this paper. We trained BERT-large using a 4x8 configuration on 32 GPUs on commodity VMs with a sequence length of 512 and batch size of 32K. We observed a throughput of 710 example/s as opposed to 700 ex/s reported by NVIDIA on DGX-1~\cite{nvidiabert}. Thus, \varuna\ is faster on low-priority VMs compared to DGX-1 with NVlink and infiniband.

\noindent
\textbf{Takeaway: } Above experiments confirm our \textit{observation 1} that pipeline parallelism is more performant than intra-layer parallelism. Even when intra-layer parallelism is limited to GPUs of a single DGX-2 server, \varuna\ is significantly faster. Thus, intra-layer parallelism should only be used when even a single layer cannot fit in one GPU.

\subsubsection{Comparison with other Pipelining architectures}

In this subsection, we compare \varuna\ with four other systems that perform pipeline partitioning: GPipe~\cite{Gpipe}, DeepSpeed~\cite{zero}, Megatron-1F1B~\cite{narayanan2021efficient} and PipeDream~\cite{pipedream}. In DeepSpeed and Megatron-1F1B, we turn off intra-layer partitioning.

\noindent
{\bf Comparison with Gpipe.}
The GPipe implementation~\cite{gpipe-source} only supports partitioning over single node. Thus, for Gpipe comparison alone, we use a BERT-72 model with 72 layers and a hidden size of 1024 that fits in a single 4-GPU node.  We partition the layers for best performance for both systems.

Table~\ref{table-bert-gpipe} shows that \varuna\ is able to deliver 15-70\% better performance than Gpipe. Note that GPipe is lot more sensitive to micro-batch size than \varuna; when the compute per micro-batch is smaller, the bubble overhead of GPipe dominates, while the pipeline schedule in \varuna\ is able to manage the compute-communication overlap better. To evaluate the impact of network conditions, we used our simulator to simulate GPipe scheduling and estimate its performance in a multi-node setting, with the calibration taken for the 8.3B model (19x3).  Under normal network latency, GPipe runs about 9\% slower than \varuna.  However, when we reduced the inter-node bandwidth, the gap between \varuna\ and GPipe widens to 38\%. 

\begin{table}
	\small
	\begin{center}
		\begin{tabular}{| l | c | c | }
			\hline
			System  &  \multicolumn{2}{c|}{Examples/s/GPU} \\
			&  \varuna & GPipe \\
			\hline
			BERT-72 ($m$=16)   & 35.9  &  21.1 \\\hline
			BERT-72 ($m$=32)   & 41.8 & 36.2  \\\hline
			Simulated 8.3B (normal network) & 0.6 & 0.55 \\\hline
			Simulated 8.3B (1.5x slower net) & 0.59 & 0.48 \\\hline
			Simulated 8.3B (2x slower net) & 0.59 & 0.426 \\
			\hline
		\end{tabular}
	\end{center}
	\caption{{\bf Comparison between \varuna\ and GPipe for a 4-stage pipeline.  $m$ refers to micro-batch size.  All configs use a mini-batch size of 8192.}}
	\label{table-bert-gpipe}
	\vspace{-0.2in}
\end{table}

\begin{table}
	\small
	\begin{center}
		\begin{tabular}{| l | c | c |c | c |}
			\hline
			System  &  \multicolumn{4}{c|}{Examples/s/GPU} \\
			&  \varuna & DeepSpeed & Megatron & PipeDream \\
			\hline
			8.3B (18x4) &  0.59 & 0.47 & 0.52  & OOM \\\hline
			2.5B (9x8)  &  1.5  & 1.24 & 1.31  & OOM \\\hline
		\end{tabular}
	\end{center}
	\caption{{\bf Comparison of \varuna, DeepSpeed, Megatron-1F1B and PipeDream for the 8.3B and 2.5B GPT-2 model on single-GPU VMs. Mini-batch size is 2400.}}
	\label{table-deepspeed}
	\vspace{-0.2in}
\end{table}

\noindent
{\bf Comparison with Deepspeed, Megatron, \& PipeDream.} 
Table~\ref{table-deepspeed} compares \varuna\ with these systems in single-GPU multi-node setting.  For this experiment, we isable intra-layer parallelism and other orthogonal optimizations like Zero for a fair comparison of pipeline efficiency of all systems.  We run 8.3B GPT-2 and 2.5 GPT-2 models on commodity 1-GPU low-pri VMs. We used a configuration of 18x4 and 9x9 for the two models, respectively. For the 8.3B/2.5B model, each of the stages had 4/6 transformer layers. Note that PipeDream, because of its storing $P$ copies of parameters (for a pipeline depth of $P$), cannot fit massive models in GPU memory, and hence is reported as OOM in the table. We also ran PipeDream-2BW but found that it did not converge (see Appendix). As shown in the table, \varuna\ performs 20-26\% better than DeepSpeed and 13-14\% better than Megatron-1F1B with higher gains as the pipeline gets longer.

\noindent
{\bf Takeaway: }
These results align with our \textit{observation 3} that \varuna\ is able to outperform other pipeline schedules since its design elements specifically cater to network latency/jitter.

\subsubsection{Scaling}

The ability of \varuna\ to scale across larger clusters can be seen from Figure~\ref{fig:megatron-8B}.  Going from 56 GPUs to 288 GPUs (5.1x more GPUs), the per-GPU performance of \varuna\ drops only by about 7.5\%. One can also see that the performance of \varuna\ in TFlops/s/GPU remains roughly the same going from a 2.5B model to a 200B model.  This demonstrates the ability of \varuna\ to scale at the same efficiency to massive models, unlike architectures like Megatron which have performance cliffs.
  
\noindent
{\bf Takeaway: } Interestingly, because \varuna\ does not require dedicated hyperclusters and can make use of opportunistic spot VMs, it can get much larger number of GPUs for a given job, at least for short periods.  Coupled with near linear scaling, \varuna\ can train models much faster (\eg, 2x faster on 2x more GPUs),  at the {\em similar dollar cost} (1000 GPUs for 2 days costs the same as 500 GPUs for 4 days in public clouds).
Thus, \varuna\ can simultaneously improve both cost and time-to-completion.

\subsection{Auto configuration and Job Morphing}

All prior systems are designed to run on a fixed number of GPUs. A unique feature of \varuna\ is its ability to adapt to dynamic spot VM availability by
quickly identifying and morphing into the best performing configuration for a job.

\medskip \noindent \textbf{Simulator:} Our auto configuration is driven by \varuna's profile-driven simulator. Table~\ref{tab-simulator} shows the predicted and actual mini-batch times on commodity VMs for several configurations. As can be seen, the simulator estimates are quite accurate, and within 5\% error margin compared to the actual values, pointing to the efficacy and sufficiency of the scale-invariant calibration of the simulator.

The time to run the simulator depends on the pipeline depth ($P$) and the number of micro-batches ($N_m$). For a 128-GPU job that uses a batch size of 8192, the simulator takes 660ms for P=36,  376ms for P=24 and 391ms for P=18, which is quick enough to react to change in spot VM availability.

\begin{table}
	\small
	\begin{center}
		\begin{tabular}{| l | c | c |c| }
			\hline
			Model & Config        &  \multicolumn{2}{c|}{Minibatch time (s)} \\
			& ($P\times D$) & Estimated & Actual \\
			\hline
			8.3B & 36x3 & 142.8 &  140.3 \\\hline
			8.3B & 36x2 & 198.7 &  201.3 \\\hline
			8.3B & 36x1 & 368.3 &  390 \\\hline
			8.3B & 24x4 & 144.7 & 149.9 \\\hline
			8.3B & 24x2 & 272.4 & 280.4 \\\hline
			8.3B & 18x6 & 139.6 & 139.1 \\\hline
			8.3B & 18x4 & 202.6 & 203.1 \\\hline
			8.3B & 18x3 & 266.6 & 263.8 \\\hline
			2.5B & 27x2  &  115.7 & 116.8 \\\hline
			2.5B & 18x3 &   92.6  &  96.6 \\\hline
			2.5B & 9x7 &   68.9  &  70.1 \\\hline
			2.5B & 6x10 &   77.5  &  75.2 \\\hline
			\hline
		\end{tabular}
	\end{center}
	\caption{{\bf Accuracy of simulator estimates for various models and configurations}}
	\label{tab-simulator}
	\vspace{-0.2in}
\end{table}

\begin{figure}
\centering{
\includegraphics[width=\linewidth]{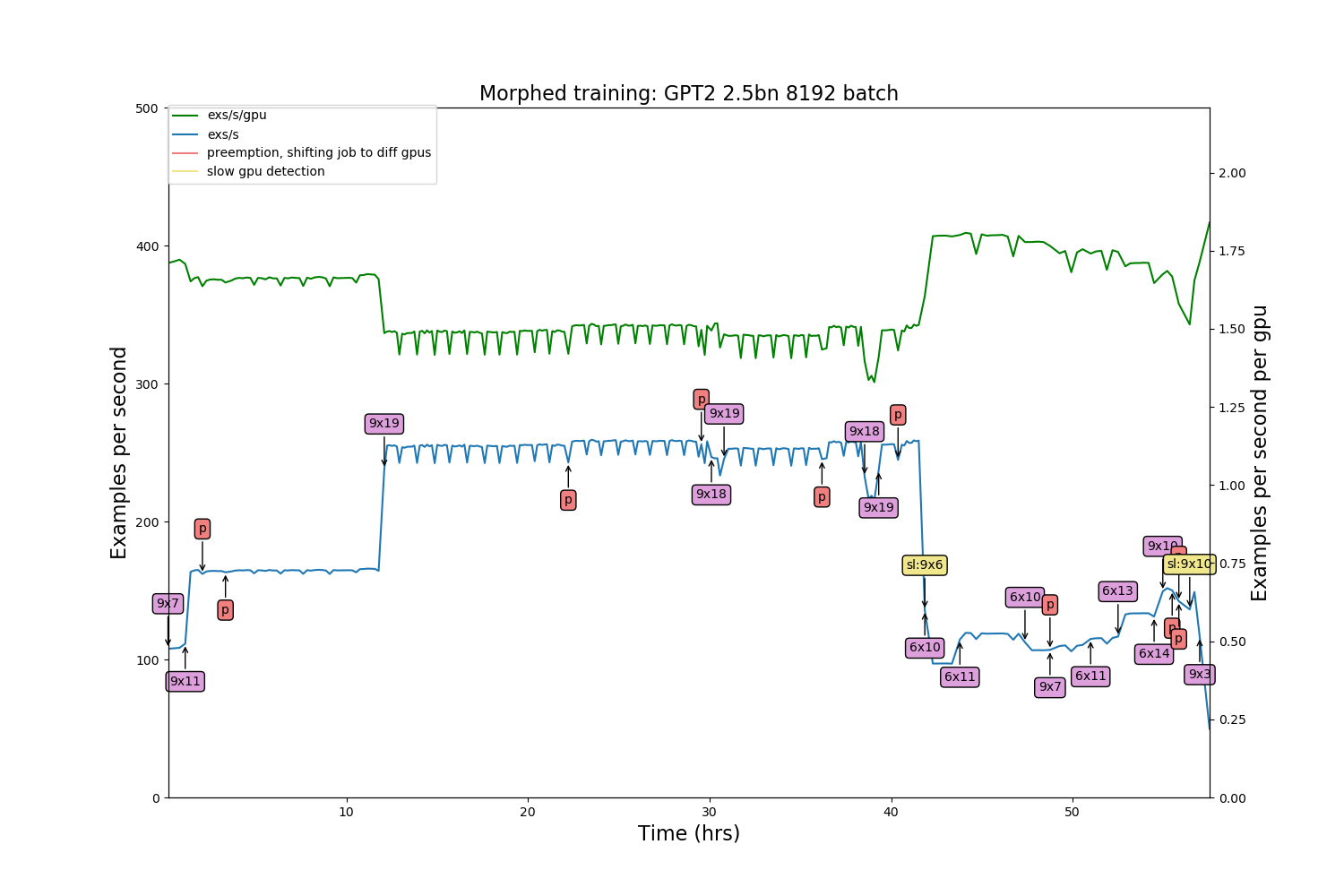}
\caption{{\bf Dynamic timeline of \varuna\ for GPT-2 2.5 billion parameter model (60 hours).}}
\label{fig:dynamic-long}
\vspace{-0.2in}
}
\end{figure}

\medskip \noindent \textbf{Morphing:} We now evaluate the robustness and flexibility of \varuna\ to adapt to variable amount of resources in the cluster, as VMs get pre-empted, and come back. For maximizing throughput, we rely on \textit{observation 4}, and request 1-GPU VMs as they are more readily available. Even though all cross-GPU communications in 1-GPU VMs go over the network, \varuna\ is able to train on 1-GPU VMs at almost the same performance as 4-GPU VMs. For example, on 72 GPUs, \varuna\ gets a throughput of 1.77 ex/s/gpu on 1-GPU VMs compared to 1.81 ex/s/GPU on 4-GPU VMs.

Figure~\ref{fig:dynamic-long} shows \varuna's training performance in examples/s and examples/s/GPU for the GPT-2 2.5 billion model over 60 hours. Each morphing event is labeled with the configuration dynamically updated by \varuna, reacting to change in available VMs. Morphing events where configuration did not change (e.g. due to replacement of a preempted machine) are marked by the letter \textit{p} while the periodic spikes are checkpointing events.  One can see that training throughput (examples/s) varies from 50 to 250 (5x) while the per-GPU performance of \varuna\ varies by only 15\% percent.

\noindent
{\bf Takeaway:} \varuna\ is able to scale effectively to utilize the dynamic spot VM available capacity, while preserving training performance.

\subsection{Accuracy and Time-to-Convergence}

\begin{figure}
	\centering{
		\includegraphics[width=0.8\linewidth]{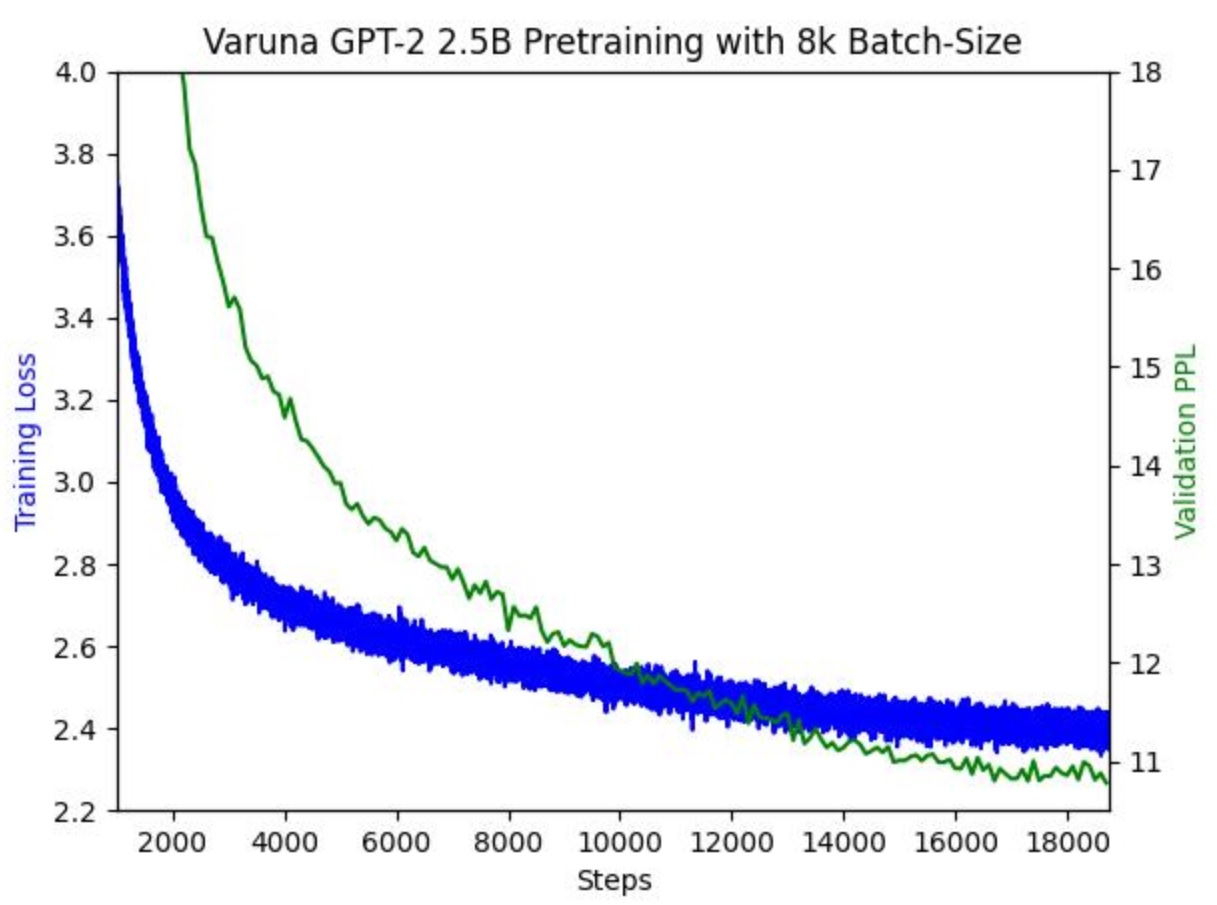}
		\caption{{\bf Convergence of GPT-2 2.5 billion model with \varuna\ using 8192 batch size}} 
		\label{fig:gpt2accuracy}
		\vspace{-0.1in}
	}
\end{figure}

\varuna\ (and all pipeline schemes in general) require mini-batch size to be about 6x larger for efficiency.  It is widely accepted that larger mini-batch training can be done as efficiently as smaller batch-sizes; \eg, for the same number of training examples, BERT-large achieves the same accuracy as mini-batch sizes vary from 512 to 65536~\cite{bert76min}.

However, to demonstrate convergence on a much larger model, we trained a 2.5 billion GPT-2 model on \varuna\ using a batch size of 8192.  Our baseline is the Megatron-2.5B model that was trained with a batch size of 512 on 300K iterations~\cite{shoeybi2019megatron}. We reduce the number of training iterations for \varuna\ by 16x to 18.75K (since we use a 16x larger batch size), thus, ensuring that
both \varuna\ and Megatron process the same number of training examples.
Figure~\ref{fig:gpt2accuracy} shows the training curve of the \varuna\ run, showing both the training loss and the validation perplexity.  \varuna\ converges to the same validation perplexity of 10.81 reported in Fig. 6 in ~\cite{shoeybi2019megatron}.  Further, we also computed WikiText103 perplexity on the trained model, and obtained a ppl of 12.78, roughly the same as the 12.76 reported in~\cite{shoeybi2019megatron}. The Lambada accuracy was 61.25\% as against 61.73\% reported, which is within noise range of the Lambada accuracy across multiple runs.  

\noindent
{\bf Takeaway:} Since \varuna\ is 4.1$\times$ faster than Megatron in examples processed per second per GPU (Figure~\ref{fig:megatron-2-5B}) and processes the same number of examples as Megatron to achieve the desired accuracy, \varuna\ improves time-to-convergence by 4.1$\times$ on commodity VMs for the 2.5B GPT-2 model.

\section{Conclusion}
\label{sec-conclude}

By opening up the possibility of training massive deep learning models with billions of parameters, on commodity VMs, \varuna\ reduces the cost of training such models by a significant factor. Further, removing the dependence on specialized hyperclusters allows such training jobs to scale to much larger numbers of GPUs opportunistically.  We believe that with this combination of low cost, high performance, and higher scale, \varuna\ can significantly accelerate the pace of innovation in large-scale AI models.

\bibliography{main}

\begin{thebibliography}{10}

\bibitem{nvidiabert}
{BERT pre-training performance}.
\newblock
  \url{https://github.com/NVIDIA/DeepLearningExamples/blob/master/PyTorch/LanguageModeling/BERT/README.md#pre-training-nvidia-dgx-1-with-16g}.

\bibitem{mxnetdl}
{Dynamic Training with Apache MXNet}.
\newblock
  \url{https://github.com/awslabs/dynamic-training-with-apache-mxnet-on-aws}.

\bibitem{gpipe-source}
{GPipe Source code}.
\newblock
  \url{https://github.com/tensorflow/lingvo/blob/master/lingvo/core/gpipe.py}.

\bibitem{APEX}
Nvidia apex library.
\newblock \url{https://github.com/NVIDIA/apex}.

\bibitem{nvlink-dgx2}
{NVIDIA DGX-2}.
\newblock \url{https://www.nvidia.com/en-in/data-center/dgx-2/}.

\bibitem{NVLAMB}
{NVLAMB Optimizer}.
\newblock
  \url{https://developer.nvidia.com/blog/pretraining-bert-with-layer-wise-adaptive-learning-rates/}.

\bibitem{elastic}
{PyTorch Elastic}.
\newblock \url{https://github.com/pytorch/elastic}.

\bibitem{Turing}
{Turing-NLG: A 17-billion-parameter language model by Microsoft}.
\newblock
  {https://www.microsoft.com/en-us/research/blog/turing-nlg-a-17-billion-parameter-language-model-by-microsoft/}.

\bibitem{tensorflow}
Mart{\'\i}n Abadi, Paul Barham, Jianmin Chen, Zhifeng Chen, Andy Davis, Jeffrey
  Dean, Matthieu Devin, Sanjay Ghemawat, Geoffrey Irving, Michael Isard, et~al.
\newblock {TensorFlow: A System for Large-Scale Machine Learning}.
\newblock In {\em 12th {USENIX} Symposium on Operating Systems Design and
  Implementation ({OSDI} 16)}, volume~16, pages 265--283. {USENIX} Association,
  2016.

\bibitem{aws-spotvm}
Amazon.
\newblock Amazon ec2 spot instances. run fault-tolerant workloads for up to
  90\% off.
\newblock In {\em https://aws.amazon.com/ec2/spot/}.

\bibitem{gpt-3}
Tom~B Brown, Benjamin Mann, Nick Ryder, Melanie Subbiah, Jared Kaplan, Prafulla
  Dhariwal, Arvind Neelakantan, Pranav Shyam, Girish Sastry, Amanda Askell,
  et~al.
\newblock Language models are few-shot learners.
\newblock {\em arXiv preprint arXiv:2005.14165}, 2020.

\bibitem{checkpointing}
Tianqi Chen, Bing Xu, Chiyuan Zhang, and Carlos Guestrin.
\newblock Training deep nets with sublinear memory cost.
\newblock {\em arXiv preprint arXiv:1604.06174}, 2016.

\bibitem{geeps}
Henggang Cui, Hao Zhang, Gregory~R Ganger, Phillip~B Gibbons, and Eric~P Xing.
\newblock Geeps: Scalable deep learning on distributed gpus with a
  gpu-specialized parameter server.
\newblock In {\em Proceedings of the Eleventh European Conference on Computer
  Systems}, pages 1--16, 2016.

\bibitem{bert-large}
Jacob Devlin, Ming-Wei Chang, Kenton Lee, and Kristina Toutanova.
\newblock Bert: Pre-training of deep bidirectional transformers for language
  understanding.
\newblock {\em arXiv preprint arXiv:1810.04805}, 2018.

\bibitem{fan2021dapple}
Shiqing Fan, Yi~Rong, Chen Meng, Zongyan Cao, Siyu Wang, Zhen Zheng, Chuan Wu,
  Guoping Long, Jun Yang, Lixue Xia, et~al.
\newblock Dapple: A pipelined data parallel approach for training large models.
\newblock In {\em Proceedings of the 26th ACM SIGPLAN Symposium on Principles
  and Practice of Parallel Programming}, pages 431--445, 2021.

\bibitem{gantt}
Henry~Laurence Gantt.
\newblock {\em Work, wages, and profits}.
\newblock Engineering Magazine Co., 1913.

\bibitem{meshtf-source}
Google.
\newblock Mesh tensorflow - model parallelism made easier.
\newblock In {\em https://github.com/tensorflow/mesh}.

\bibitem{meshtf-talk}
Google.
\newblock Mesh-tensorflow: Model parallelism for supercomputers (tf dev summit
  ‘19).
\newblock In {\em https://www.youtube.com/watch?v=HgGyWS40g-g}.

\bibitem{harlap2017proteus}
Aaron Harlap, Alexey Tumanov, Andrew Chung, Gregory~R Ganger, and Phillip~B
  Gibbons.
\newblock Proteus: agile ml elasticity through tiered reliability in dynamic
  resource markets.
\newblock In {\em Proceedings of the Twelfth European Conference on Computer
  Systems}, pages 589--604, 2017.

\bibitem{Gpipe}
Yanping Huang, Youlong Cheng, Ankur Bapna, Orhan Firat, Dehao Chen, Mia Chen,
  HyoukJoong Lee, Jiquan Ngiam, Quoc~V Le, Yonghui Wu, et~al.
\newblock Gpipe: Efficient training of giant neural networks using pipeline
  parallelism.
\newblock In {\em Advances in Neural Information Processing Systems}, pages
  103--112, 2019.

\bibitem{checkmate}
Paras Jain, Ajay Jain, Aniruddha Nrusimha, Amir Gholami, Pieter Abbeel, Kurt
  Keutzer, Ion Stoica, and Joseph~E Gonzalez.
\newblock Checkmate: Breaking the memory wall with optimal tensor
  rematerialization.
\newblock {\em arXiv preprint arXiv:1910.02653}, 2019.

\bibitem{cpumemoverhead}
Chen Meng, Minmin Sun, Jun Yang, Minghui Qiu, and Yang Gu.
\newblock Training deeper models by gpu memory optimization on tensorflow.
\newblock In {\em Proc. of ML Systems Workshop in NIPS}, 2017.

\bibitem{azure-lowprivm}
Microsoft.
\newblock Use low-priority azure vms with batch.
\newblock In {\em
  https://docs.microsoft.com/en-us/azure/batch/batch-low-pri-vms}.

\bibitem{pipedream}
Deepak Narayanan, Aaron Harlap, Amar Phanishayee, Vivek Seshadri, Nikhil~R
  Devanur, Gregory~R Ganger, Phillip~B Gibbons, and Matei Zaharia.
\newblock Pipedream: generalized pipeline parallelism for dnn training.
\newblock In {\em Proceedings of the 27th ACM Symposium on Operating Systems
  Principles}, pages 1--15, 2019.

\bibitem{narayanan2020memory}
Deepak Narayanan, Amar Phanishayee, Kaiyu Shi, Xie Chen, and Matei Zaharia.
\newblock Memory-efficient pipeline-parallel dnn training.
\newblock {\em arXiv preprint arXiv:2006.09503}, 2020.

\bibitem{narayanan2021efficient}
Deepak Narayanan, Mohammad Shoeybi, Jared Casper, Patrick LeGresley, Mostofa
  Patwary, Vijay Korthikanti, Dmitri Vainbrand, Prethvi Kashinkunti, Julie
  Bernauer, Bryan Catanzaro, et~al.
\newblock Efficient large-scale language model training on gpu clusters.
\newblock {\em arXiv preprint arXiv:2104.04473}, 2021.

\bibitem{or2020resource}
Andrew Or, Haoyu Zhang, and Michael Freedman.
\newblock Resource elasticity in distributed deep learning.
\newblock {\em Proceedings of Machine Learning and Systems}, 2:400--411, 2020.

\bibitem{pytorch}
Adam Paszke, Sam Gross, Soumith Chintala, and Gregory Chanan.
\newblock Pytorch, 2017.

\bibitem{allreduce}
Pitch Patarasuk and Xin Yuan.
\newblock Bandwidth optimal all-reduce algorithms for clusters of workstations.
\newblock {\em Journal of Parallel and Distributed Computing}, 69(2):117--124,
  2009.

\bibitem{gpt-2}
Alec Radford, Jeffrey Wu, Rewon Child, David Luan, Dario Amodei, and Ilya
  Sutskever.
\newblock Language models are unsupervised multitask learners.
\newblock {\em OpenAI Blog}, 1(8):9, 2019.

\bibitem{zero}
Samyam Rajbhandari, Jeff Rasley, Olatunji Ruwase, and Yuxiong He.
\newblock {ZeRO: Memory Optimization Towards Training A Trillion Parameter
  Models}.
\newblock {\em arXiv preprint arXiv:1910.02054}, 2019.

\bibitem{zero-infinity}
Samyam Rajbhandari, Olatunji Ruwase, Jeff Rasley, Shaden Smith, and Yuxiong He.
\newblock Zero-infinity: Breaking the gpu memory wall for extreme scale deep
  learning.
\newblock {\em arXiv preprint arXiv:2104.07857}, 2021.

\bibitem{trainingcost}
Or~Sharir, Barak Peleg, and Yoav Shoham.
\newblock The cost of training nlp models: A concise overview.
\newblock {\em arXiv preprint arXiv:2004.08900}, 2020.

\bibitem{meshtf}
Noam Shazeer, Youlong Cheng, Niki Parmar, Dustin Tran, Ashish Vaswani, Penporn
  Koanantakool, Peter Hawkins, HyoukJoong Lee, Mingsheng Hong, Cliff Young,
  et~al.
\newblock Mesh-tensorflow: Deep learning for supercomputers.
\newblock In {\em Advances in Neural Information Processing Systems}, pages
  10414--10423, 2018.

\bibitem{shoeybi2019megatron}
Mohammad Shoeybi, Mostofa Patwary, Raul Puri, Patrick LeGresley, Jared Casper,
  and Bryan Catanzaro.
\newblock Megatron-lm: Training multi-billion parameter language models using
  gpu model parallelism.
\newblock {\em arXiv preprint arXiv:1909.08053}, 2019.

\bibitem{sivathanu2019astra}
Muthian Sivathanu, Tapan Chugh, Sanjay~S. Singapuram, and Lidong Zhou.
\newblock Astra: Exploiting predictability to optimize deep learning.
\newblock In {\em Proceedings of the Twenty-Fourth International Conference on
  Architectural Support for Programming Languages and Operating Systems},
  ASPLOS '19, pages 909--923, New York, NY, USA, 2019. ACM.

\bibitem{Transformers}
Ashish Vaswani, Noam Shazeer, Niki Parmar, Jakob Uszkoreit, Llion Jones,
  Aidan~N Gomez, {\L}ukasz Kaiser, and Illia Polosukhin.
\newblock Attention is all you need.
\newblock In {\em Advances in neural information processing systems}, pages
  5998--6008, 2017.

\bibitem{wagenlander2020spotnik}
Marcel Wagenl{\"a}nder, Luo Mai, Guo Li, and Peter Pietzuch.
\newblock Spotnik: Designing distributed machine learning for transient cloud
  resources.
\newblock In {\em 12th $\{$USENIX$\}$ Workshop on Hot Topics in Cloud Computing
  (HotCloud 20)}, 2020.

\bibitem{wang2019characterizing}
Mengdi Wang, Chen Meng, Guoping Long, Chuan Wu, Jun Yang, Wei Lin, and Yangqing
  Jia.
\newblock Characterizing deep learning training workloads on alibaba-pai.
\newblock In {\em 2019 IEEE International Symposium on Workload
  Characterization (IISWC)}, pages 189--202. IEEE, 2019.

\bibitem{xiao2018gandiva}
Wencong Xiao, Romil Bhardwaj, Ramachandran Ramjee, Muthian Sivathanu, Nipun
  Kwatra, Zhenhua Han, Pratyush Patel, Xuan Peng, Hanyu Zhao, Quanlu Zhang,
  et~al.
\newblock Gandiva: Introspective cluster scheduling for deep learning.
\newblock In {\em 13th $\{$USENIX$\}$ Symposium on Operating Systems Design and
  Implementation ($\{$OSDI$\}$ 18)}, pages 595--610, 2018.

\bibitem{pipemare}
Bowen Yang, Jian Zhang, Jonathan Li, Christopher R{\'e}, Christopher~R Aberger,
  and Christopher De~Sa.
\newblock Pipemare: Asynchronous pipeline parallel dnn training.
\newblock {\em arXiv preprint arXiv:1910.05124}, 2019.

\bibitem{bert76min}
Yang You, Jing Li, Sashank Reddi, Jonathan Hseu, Sanjiv Kumar, Srinadh
  Bhojanapalli, Xiaodan Song, James Demmel, Kurt Keutzer, and Cho-Jui Hsieh.
\newblock Large batch optimization for deep learning: Training bert in 76
  minutes.
\newblock In {\em International Conference on Learning Representations}, 2019.

\end{thebibliography}
\bibliographystyle{plain}
\section{Appendix}

\subsection{PipeDream-2BW Convergence}
PipeDream-2BW boosts pipeline training throughput by sacrificing sync SGD semantics, which can lead to sub-optimal accuracy results.  Although the authors show convergence results for Bert, asynchronous training is not a common practice today, primarily since it is unclear how final accuracy will be affected due to stale backward updates~\cite{wang2019characterizing}.

\begin{figure}[h]
	\begin{center}
		\includegraphics [width=1.0\linewidth]{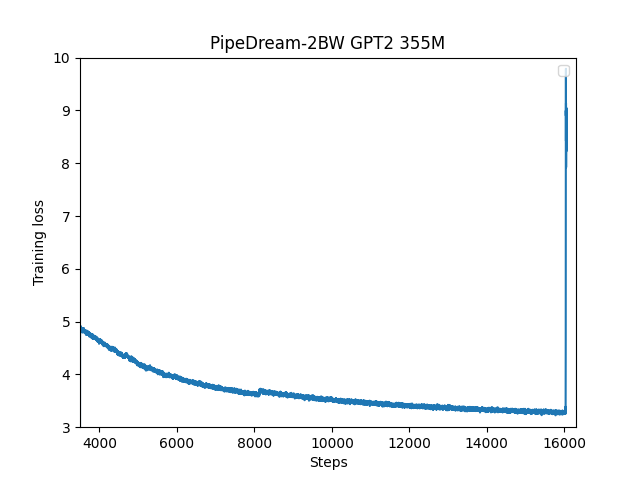}
		\caption{PipeDream-2BW Loss Curve for 355M-parameter GPT2 model}
		\label{fig:pipedream_loss}
	\end{center}
\end{figure}

For 355M GPT2 model, the authors report a WikiText PPL of 19.56 as opposed to 19.28 reported by Nvidia~\cite{shoeybi2019megatron} and 19.16 PPL with Varuna using 8k batch size (lower is better). We attempted to train a 355M GPT2 model from scratch using PipeDream-2BW in order to compute Lambada accuracy, another standard benchmark for this dataset. However, 
we found that, after 16k iterations, the training diverged and the loss shot up as shown in Figure~\ref{fig:pipedream_loss}. We used the same set of hyperparameters as listed in the PipeDream-2BW paper, a batch size of 512, Adam optimizer with learning rate of \ $10^{-4}$ \, with initial warmup and subsequent linear decay and a maximum sequence length of 512 with 6 pipeline stages on 48 GPUs.

\end{document}